\documentclass[12pt]{article}
\usepackage{amsmath}
\usepackage{graphicx}
\usepackage{enumerate}
\usepackage{natbib}
\usepackage{url}

\usepackage{amsfonts}
\usepackage{dsfont}
\usepackage{amsmath}
\usepackage{float}
\usepackage[font=small,labelfont=bf]{caption}
\usepackage{setspace}
\usepackage{booktabs}
\usepackage{caption}
\usepackage{subcaption}
\usepackage{xcolor}
\usepackage{parskip}
\usepackage{authblk}
\setcitestyle{authoryear,open={(},close={)}}

\renewcommand{\P}{\mathbb{P}}

\begin{document}

\def\spacingset#1{\renewcommand{\baselinestretch}%
{#1}\small\normalsize} \spacingset{1}

\title{\bf  Zombie Epidemic --- on Modeling the Effect of Interventions}
\author[1]{Kaisa Ek}
\author[1]{Aleksi Avela}
\author[2]{Willehard Haaki}
\author[3]{Sami Helander}
\author[4]{Ville Lumme}
\author[1]{Terho Mutikainen}
\author[1]{Jaakko Pere}
\author[1]{Natalia Vesselinova}
\author[3]{Lauri Viitasaari}
\author[1]{Pauliina Ilmonen}
\affil[1]{Department of Mathematics and Systems Analysis, Aalto University
School of Science, Espoo, Finland, \{firstname.lastname@aalto.fi\}}
\affil[2]{Department of Psychiatry, University of Turku, Turku, Finland,
hawker@utu.fi}
\affil[3]{Department of Information and Service Management,
Aalto University School of Business, Espoo, Finland,
\{firstname.lastname@aalto.fi\}}
\affil[4]{Department of Architecture, Aalto University School of Arts,
Design and  Architecture, Espoo, Finland, ville.lumme@aalto.fi}

\maketitle

\bigskip
\begin{abstract}
The recent COVID-19 pandemic has highlighted the need of studying extreme,
life-threatening phenomena in advance. In this article, a zombie epidemic in
Uusimaa region in Finland is modeled. A stochastic agent based simulation model
is proposed and extensive simulations are conducted for this purpose. The model
utilizes knowledge on defensive human behavior during crises. Studying the
effects of a hypothetical zombie attack resembles examining the spread of deadly
diseases and of rumors. A zombie attack is simulated in the most densely
populated region in Finland. The region's exact population densities over its
rasterized geographical map are utilized. Furthermore, the simulations are used
to study the effect of implementing a (strict or partial) quarantine area in the
epicenter. Computationally efficient Scala codes and video animations of the
simulated epidemics are provided. The main findings emphasize the importance of
implementing very strict measures, without delay, to stop the outbreak.
\end{abstract}


\section{Introduction}
\label{sec:intro}

The recent COVID-19 pandemic has underlined the need to study hypothetical
scenarios such as a zombie apocalypse as a means to prepare for real-world
disasters and pandemics. Investigating such extreme scenarios can enhance our
understanding of what measures can prove effective, and hence, can contribute to
the development of emergency response strategies, better psychological
resilience, and overall societal preparedness.

A zombie infection has many similarities with more ordinary diseases, the spread
of which is often modeled with compartmental models such as the
susceptible-infectious-recovered (SIR) model. Compartmental models are used to
model, e.g., the spread of COVID-19 \citep{cooper2020}, influenza A
\citep{casagrandi2006} and Ebola \citep{berge2017}. The work of \cite{munz2009}
is one of the earliest examples of modeling zombie attacks with a compartmental
model called the susceptible-zombie-removed (SZR) model. As the SIR model,
simple SZR model can be expressed in terms of a system of ordinary differential
equations (ODEs).

The estimation of the SZR model parameters is challenging due to the apparent
lack of data since --- to this day --- there has not been reports of zombie
attacks. Nevertheless, some educated guesses for model parameters, such as the
``biting'' rate of the zombies, have been made. \cite{witkowski2013} use
Bayesian estimation of the model parameters based on data given by popular
culture zombie films. \cite{mcgahan2021} fit and test the predictive power of
multiple models based on data from the game ``Humans versus Zombies'' played by
real humans at Utah State University.

ODEs of the SZR model involve sizes of the populations, whereas in the
traditional SIR model frequencies are usually used. This has a large effect on
the SZR model in heterogeneous populations\footnote{The interested reader is
referred to \cite{alemi2015} for further details about the differences between
the traditional SIR and SZR models.}. Also, in a zombie attack, an active
counterattack from the susceptible population is required to defeat the zombies.
Therefore, zombie models could be used in various contexts, such as in the
spread of rumors \citep{amaral2018, amaral2020} or HIV \citep{libal2023}, where
counteraction from the ``healthy population'' is required to kill the
``infection''.

We provide a simulation model, that is motivated by the SZR model. The SZR model
is characterized by the rate for zombies infecting humans and the rate for
humans killing zombies. In turn, we use the probabilities of these events
(Section~\ref{sec:behavior} and Section~\ref{sec:interactions}). Simulations by
\cite{alemi2015} incorporated spatial structure into the SZR model. In our
simulations, we use real population density data (Section~\ref{sec:map}) and
zombies and humans walk randomly in the corresponding lattice
(Section~\ref{sec:movement}). \cite{munz2009, mendona2019, alemi2015} augmented
the usual classes $S$, $Z$ and $R$ of the SZR model in various ways. In our
model, the transition from human to zombie upon infection is not instant, but
there is an incubation period (Section~\ref{sec:agents}); the susceptible
population is divided into subpopulations based on human defensive behaviors
(Section~\ref{sec:behavior}); and the possibility of building a quarantine zone
around the epicenter is introduced (Section~\ref{sec:scenarios}). Our
contribution can be summarized in: we provide a computationally efficient agent
based stochastic zombie epidemic simulator that models a real geographical area
together with its detailed (1km-by-1km) population densities and incorporates
human behavior typical for extreme situations as described by earlier scientific
research. Furthermore, the simulations are used to study the effect of
implementing a (strict or partial) quarantine. The large scale simulations can
be run on a regular laptop. Scala codes and map animation videos of the
simulated epidemics are freely available in the project's GitHub repository
\citep{repo}.

The rest of this article is organized as follows. The model is described in
Section~\ref{sec:model}. The results from the simulations, with and without
implementing a quarantine zone, are presented in Section~\ref{sec:simulations}.
The results are discussed in Section~\ref{sec:discussion}, and ideas for
extending the study are provided in Section~\ref{sec:future}.

\section{Zombie simulation model}
\label{sec:model}

\subsection{Human defensive behavior}
\label{sec:behavior}

Human defensive behaviors are typically categorized into Freeze
(hypervigilance), Flight (escaping), Fight, Fright (Freeze with tonic immobility
or Flop with pliable immobility) and Faint (syncope). If communication with the
perceived threat is possible, another possible response is Fawn, characterized
by submission to the situation. In mammals, the typical sequence begins with
Freeze, allowing them to assess the situation and to avoid detection by
remaining still; followed by Flight, if escaping offers a better chance of
survival; and finally, if escape is not an option, Fight. When these options are
not viable, the response may escalate to Fright, resulting in tonic or pliable
immobility, similar to an opossum playing dead or, tragically, as seen in some
human victims of sexual violence.  The response Faint is not present in most
mammals, but it may occur as a human response.

In a zombie-type scenario, it makes sense to classify the described human
reactions into three distinct actions. First, active escape (Flight), which
involves running away and hiding; second, aggression (Fight), which includes
various forms of attacking; and third, passive submission (Freeze) which
encompasses freezing, fainting, or flopping --- where the individual takes no
action in saving themselves. Determining the precise probabilities of different
reactions is quite challenging.

In \cite{blanchard2001human}, study participants read a set of scenarios
involving threatening situations, and chose their most likely primary defensive
response to each. In the study, most male and female responses were highly
correlated, with the exception of the response yell, scream or call for help.
The proportion of females who reported choosing to flee grew as the threat
became more unambiguous.  The proportion of females who reported choosing to
attack also increased, but at a slower rate. A limitation of the study is that
it relies on self-assessment, making it difficult to adequately assess the
freeze response \citep{krupic2017situational}. In \cite{schmidt2008exploring},
when study participants were exposed to inhaling CO$_2$ enriched air,  $18\%$ of
them reported mild feeling of freezing (tonic immobility), $13\%$ of them
reported significant feeling of freezing, $19\%$ of them reported mild desire to
flee, and $20\%$ reported modest or greater desire to flee.

Considering that our study assumes a direct unambiguous threat (zombies) with no
possibility of communication, and presumes a typical urban environment with
usual opportunities to escape and shelter, most subjects would flee. An attack
response is also possible, but as in \cite{blanchard2001human}, it would be a
less common. Since participants in \cite{schmidt2008exploring} volunteered for
the study and they were aware that they would face a threatening situation, the
proportion of involuntary freeze responses during a zombie attack could be
approximated to be higher than the proportion of study subjects who reported
significant feelings of freezing in \cite{schmidt2008exploring}. Moreover, as
running away and hiding does not help when exposed to inhaling CO$_2$ enriched
air in laboratory settings, the proportion of humans fleeing in a zombie attack
could be approximated to be higher than the proportion of the study subjects
that reported desire to flee in \cite{schmidt2008exploring}. Therefore, we
estimate the probabilities in our scenario to be $20\%$ freezing, $55\%$ fleeing
and $25\%$ attacking.

\subsection{The Simulation Setting}
\label{sec:setting}

\begin{figure}
 	\centering
 	\begin{subfigure}{.49\linewidth}
 		\centering
 		\includegraphics[width=.59\linewidth]{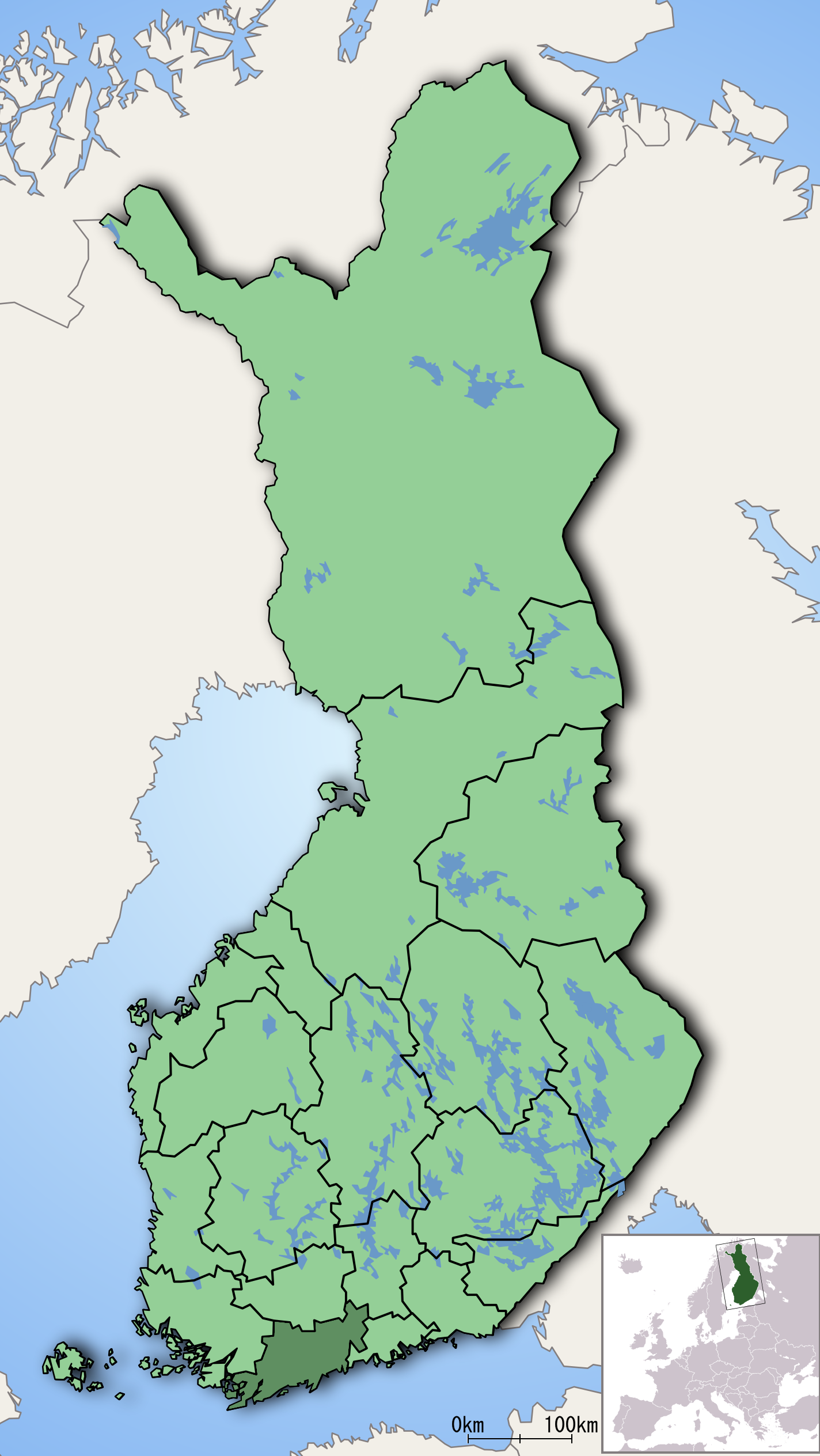}
 	\end{subfigure}~
 	\begin{subfigure}{.49\linewidth}
 		\centering
 		\includegraphics[width=\linewidth]{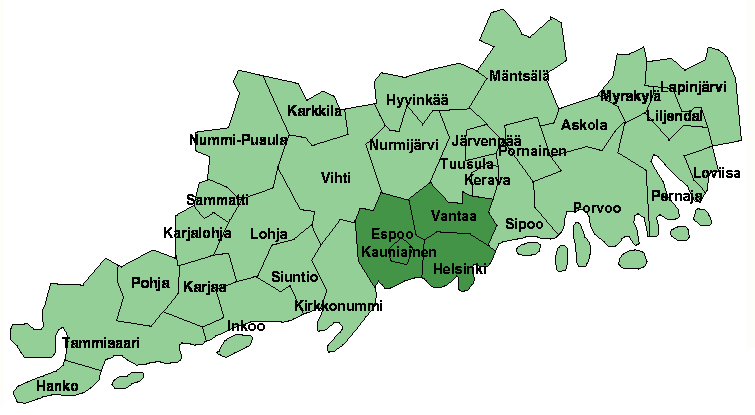}
 	\end{subfigure}\\[1mm]
 	\caption{(left) A map of Finland and its location in Europe \citep{Finland}.
 	(right) A map of Uusimaa region \citep{Uusimaa}, the most densely populated
 	area in Finland. Helsinki region (Helsinki, Espoo, Vantaa and Kauniainen) is
 	highlighted in dark green.}
 	\label{fig:map}
 \end{figure}

We place the zombie epidemic to Uusimaa region in southern Finland
(Figure~\ref{fig:map}), and we use the true shape, size and local population
densities of the region. Uusimaa is the most densely populated area in Finland
covering approximately $3\%$ of the country's total area, but accounting for
more than $30\%$ of its population. Furthermore, it serves as the center of
governmental power, including Helsinki and the capital area. This introduces an
additional challenge, as in crisis situations, it is the governmental entities
that make decisions and lead the emergency responses.

\subsubsection{Map}
\label{sec:map}

A map for the simulations is formed by rasterizing the Uusimaa region
(Figure~\ref{fig:map}) into a 1km-by-1km grid of squares.  Population of
$1\,704\,456$ people, the true population size of Uusimaa, is placed on the map.
The distribution of the population follows the true population density
distribution over Uusimaa region as recorded in year 2022 \citep{geo},
discretized into the same grid of squares used in constructing the map. During
the simulations, agents (see Section~\ref{sec:agents}) move and interact with
other agents. The regions outside of Uusimaa, as well as larger bodies of water
(such as lakes and the seashore) are impassable. At the beginning of each
simulation run, one single zombie is placed in the center of Helsinki.

\subsubsection{Agents}
\label{sec:agents}

A simulation has two types of agents --- humans and zombies --- who move and
interact with each other. Zombies are represented as dead if they are defeated
in a fight by a human. Humans that are defeated by zombies transform into
zombies themselves.

\subsubsection{Interactions}
\label{sec:interactions}

During each one-hour simulation step, every zombie interacts with every human
within the same 1km-by-1km square.

First, a reaction for the human is selected randomly from Fight, Flight or
Freeze, with probabilities $\P(Fight)=0.25$,  $\P(Flight)=0.55$, and
$\P(Freeze)=0.20$ (see Section~\ref{sec:behavior}). Then, depending on the
randomly chosen reaction, the interaction proceeds as follows:
\begin{itemize}
	\item Fight $\rightarrow$ The human has the probability of $\P(Win)=0.5$ to
	defeat the zombie, and the probability $\P(Lose)=1-\P(Win)$ to lose and to
	be transformed into a zombie.
	\item Flight $\rightarrow$ The human has the probability of
 	$\P(Escape)=0.70$ of successfully escaping, and the probability
 	$\P(Caught)=0.30$ of being caught. If caught, the human is forced to fight,
 	with the probability of only $\P(Win)=0.10$ for defeating the zombie.
	\item Freeze $\rightarrow$ The human, immobilized by psychological stress,
	is compelled to fight the zombie with the probability of only $\P(Win)=0.05$
	for defeating it.
\end{itemize}

In every encounter where the human wins a fight, the zombie is killed. However,
if a human successfully flees, then the interaction does not lead to a fight.
Thus the probability for a zombie to die in a random interaction is
$
0.25\cdot 0.5+0.55\cdot 0.3\cdot 0.1 + 0.2\cdot 0.05 = 0.1515.
$
Dead zombies neither move nor interact with any other agents. In every encounter
where the zombie wins a fight, the human is transformed into a zombie after an
incubation period of 1 time step. The probability for a human transforming into
a zombie in a random interaction is $ 0.25\cdot (1-0.5)+0.55\cdot 0.3\cdot
(1-0.1) + 0.2\cdot (1-0.05) = 0.4635.$ Newly infected humans do not interact
with any other agents during the same time step that they are infected, but they
can still move according to the zombie movement logic.

\subsubsection{Movement}
\label{sec:movement}

Once the interactions have been resolved, zombies and humans move on the map.

Zombies move one step on the grid to any of the eight adjacent squares (up,
down, left, right, and diagonally), chosen uniformly at random from the
available ones. Impassable squares, such as map borders or water, are
disregarded. Consequently, zombies move approximately one or one and a half
($\sqrt{2}$) kilometers per an hour.
	
Humans move a total of two steps, executed one by one. Thus, the humans move
between zero to approximately three ($2\sqrt{2}$) kilometers per an hour. Humans
have an increasing tendency, depending on their distance from their home square
(that is, the square where they are at the beginning of the simulation run), to
prefer moving to a square in the three-square-wide sector towards their home
square. Again, impassable squares are disregarded. The humans' movement logic is
implemented as follows. In their home square, the humans choose a movement
direction uniformly at random from the available adjacent squares. Otherwise,
the probabilities for a human to move to the adjacent squares are as follows:
\begin{itemize}
	\item The probability to move to any square in the three square wide sector
	towards the home square is given by
	$
		p_H = \frac{3 \cdot 0.125 + x\cdot 0.05}{h},
	$
	where $h\in\{1,2,3\}$ is the number of available squares in the sector
	towards home, and $x\in\mathbb{N}$, $x\leq 12$,  is the Euclidean distance
	from the current position to the home square, rounded down to the closest
	natural number. When the distance $x\geq 13$ from home, $p_H=\frac{1}{h}$
	provided that $h>0$.
		
	\item The probability of moving to any square \textit{outside} the sector
	towards the home square (the sector away from home) is given by
	$
		p_A = \frac{5 \cdot 0.125 - x\cdot 0.05}{a},
	$
	where $a\in\{1,2,3,4,5\}$ is the number of available adjacent squares that
	do not belong to the sector towards home, and  as above, $x\in\mathbb{N}$,
	$x\leq 12$, is the distance from the current position to the home square.
	When the distance $x\geq13$ from home, $p_A=0$ provided that $h>0$.
		
	\item If no squares are available in the sector towards the home square
	(i.e., $h=0$), or if no squares are available in the sector away from home
	(i.e.,~ $a=0$), then humans move uniformly at random to any of the available
	squares.
\end{itemize}
The sector towards the home square consists of three (out of eight) squares that
are adjacent to the current position square. The ``central'' square of the
sector is the one with the smallest Euclidean distance from the home square, and
it is accompanied by the two squares that the current position square and share
an edge with the central square.

\subsection{Simulated scenarios}
\label{sec:scenarios}

The simulation study consists of three different scenarios, each simulated
$n=1\,000$ times. In the base Scenario 1, simulations are run as described
above, starting with one zombie in the center of Helsinki and no special
interventions. Scenarios 2 and 3 introduce an intervention in the form of a
quarantine area containing the Helsinki region (comprising the capital city of
Helsinki as well as the cities of Espoo, Vantaa, and Kauniainen;
see~Figure~\ref{fig:map}). In these simulations, the quarantine becomes
operational at time step $14$. This delay is intended to reflect the time
required for detecting the epidemic, coordinating a response, and setting up the
quarantine area. The specific 14-hour intervention delay was also guided by the
results from Scenario 1; see
Figure~\ref{fig:scenario-1-edge-of-helsinki-time-distribution}, which displays
the distribution of the time steps for the first zombie to reach the border of
the capital region. Our objective is to compare the spread of rapidly
propagating diseases with and without interventions. Therefore,  the quarantine
was set to begin at time step 14 to ensure it would be in place just in time for
the majority of the simulation runs, but allowing a few of the fastest zombies
to pass through in some of the runs. This approach aims to create some scenarios
where the intervention is timely, but also some scenarios where it arrives too
late.

In Scenario 2, the quarantine is strictly enforced, preventing any agent from
crossing the quarantine area border. Since distinguishing between an infected
human in their incubation period and a healthy human is challenging, all agents
attempting to pass through the border must be stopped to protect the population
outside. Therefore, humans attempting to cross the border of the quarantine area
from the Helsinki region are stopped, while zombies attempting to move from the
quarantine area to outside of the Helsinki region are killed. It is assumed that
humans can understand and follow instructions, whereas zombies cannot. All
agents attempting to  enter the quarantine area from outside of the Helsinki
region are stopped but not killed.

In Scenario 3, the quarantine area border is not fully secure to reflect the
challenges of establishing a strict quarantine in a short time over a large
area, especially given the large number of people in the Helsinki region
(approximately 1.2 million) trying to escape the outbreak. Consequently, the
quarantine area border is not completely impermeable, allowing the agents to
cross from the Helsinki region with a probability of $\P(Leak) = 0.001$. Humans
who attempt but fail to cross the quarantine border from the Helsinki region are
stopped, while zombies who try and fail to cross from inside the quarantine area
are killed. No agent is able to enter the quarantine zone from outside of the
Helsinki region.

\section{Simulation results}
\label{sec:simulations}

\begin{figure}[!th]
	\centering
	\includegraphics[width=\linewidth]{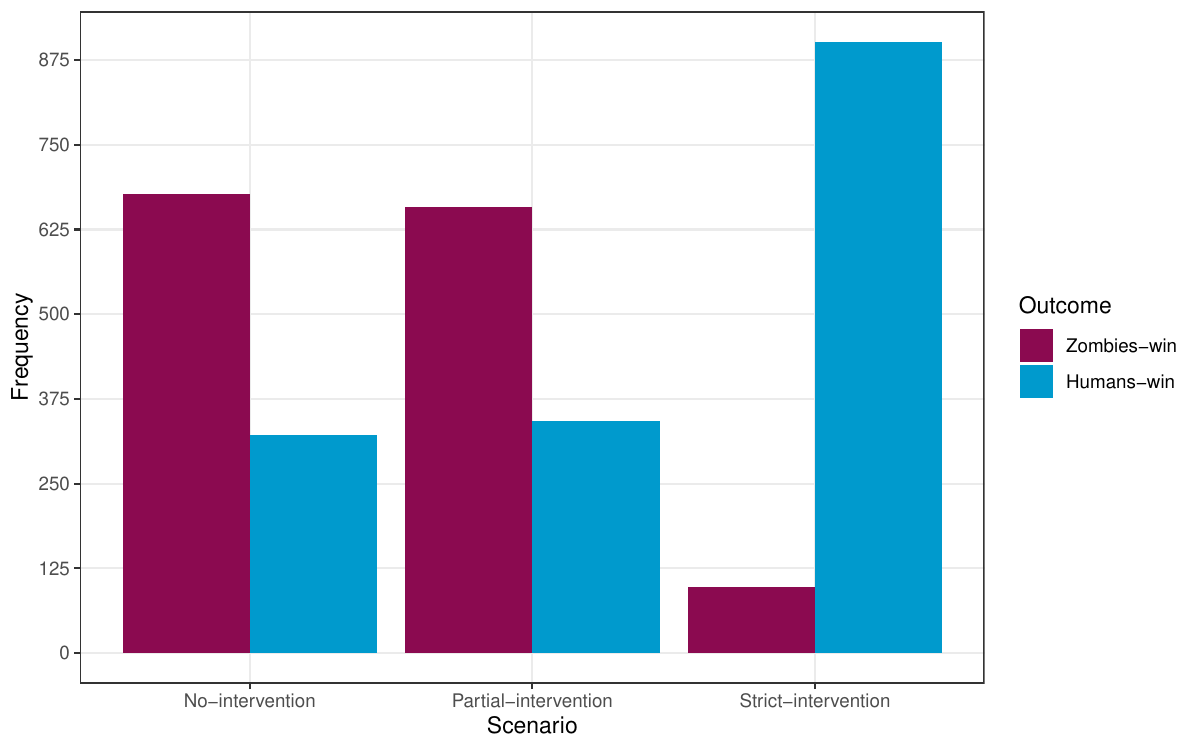}
	\caption{Number of respective wins for humans (in blue) and for zombies (in red) in all three simulated scenarios.}
	\label{fig:all-outcomes}
\end{figure}

A simulation run stops when either humans or zombies win. Humans win if all the
zombies are defeated, or in Scenario 2, also if the strict intervention has
taken place and there are no more zombies left outside of the quarantine area.
Zombies win if all living agents have transformed into zombies. The number of
wins for humans and for zombies across the three scenarios are compared in
Figure~\ref{fig:all-outcomes}. The difference between stringent measures, less
strict measures, and no intervention is evident.

In Scenario 1, which has no interventions, zombies win in $678$ out of the
$1\,000$ simulation runs. Humans win in the remaining $322$ runs. Assuming that
zombies always win if their population reaches five, the probability of a zombie
victory is $69.1\%$ which is in line with the simulation results. This
probability is calculated using probability trees and solving the corresponding
system of linear equations. The assumption that zombies always win with a
population of five is reasonable, given that the probability of five individual
zombies  being defeated in their next forthcoming fights is $0.246341^5 =
0.09\%$, where $0.246341 = \frac{0.1515}{0.1515+0.4635}$ equals the probability
of a zombie being defeated conditioned on that a fight occurs.

In Scenario 2, with strict quarantine area built at time step 14, humans win in
902 out of $1\,000$ simulation runs, while zombies win only 98 times. Strict
intervention is very effective. In the 98 zombie victories, the intervention
arrived too late illustrating that just a few escaping zombies can cause the
epidemic to start spreading uncontrollably in the areas outside of the
quarantine region.

In Scenario 3, for any zombie who tries to cross the border from the quarantine
area, the probability for being killed is $99.9\%$, and the probability for
successful escape is only $0.1\%$. However, zombies still win in 658 out of
$1\,000$  simulation runs, while humans win only in 342. This is plausible
since, if the Helsinki region is lost to zombies, the number of zombies that
eventually cross the border follows a binomial distribution with approximately
1.2 million trials (zombies inside the quarantine area) and a success
probability of $0.001$. As expected, the average time for the zombies to take
over Uusimaa is longer when a partial intervention is in place compared to no
intervention, see
Figures~\ref{fig:scenario-1-hours-to-win-zombies},~\ref{fig:scenario-1-all-runs},~\ref{fig:scenario-3-hours-to-win-zombies}
and \ref{fig:scenario-3-all-runs}.

\begin{figure}[!h]
	\centering
	\begin{subfigure}{.49\linewidth}
		\centering
		\includegraphics[width=\linewidth]{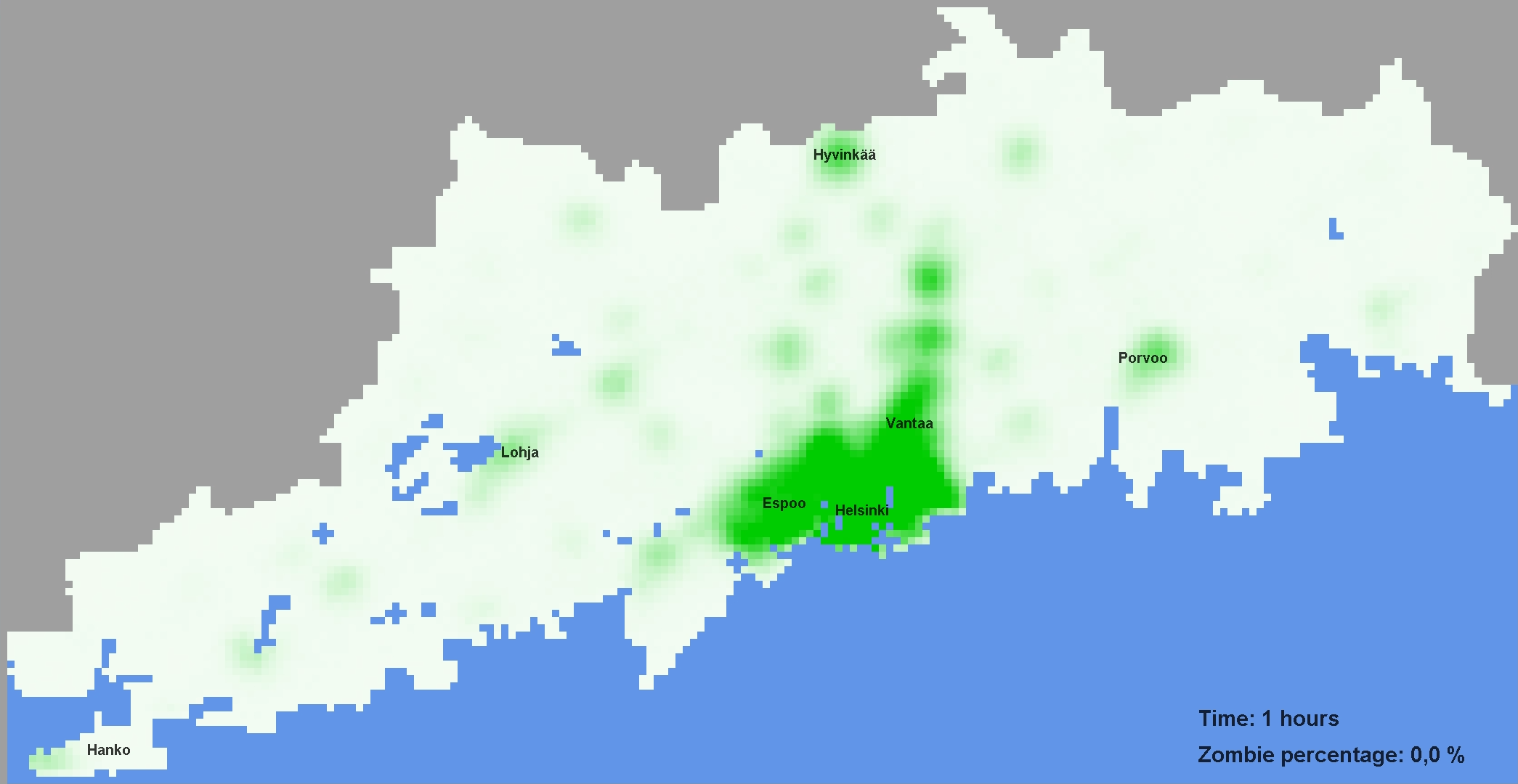}
	\end{subfigure}~
	\begin{subfigure}{.49\linewidth}
		\centering
		\includegraphics[width=\linewidth]{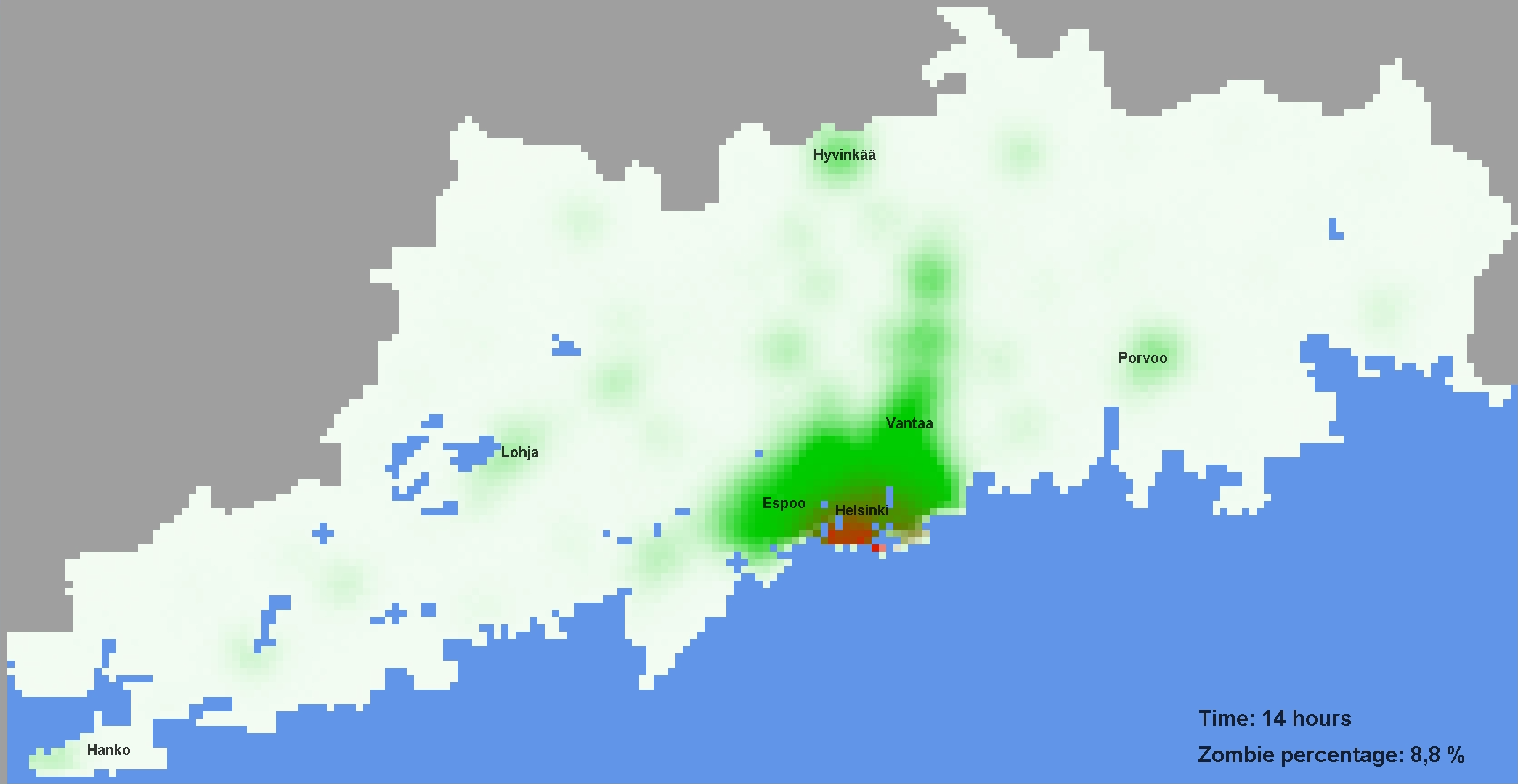}
	\end{subfigure}\\[1mm]
	\begin{subfigure}{.49\linewidth}
		\centering
		\includegraphics[width=\linewidth]{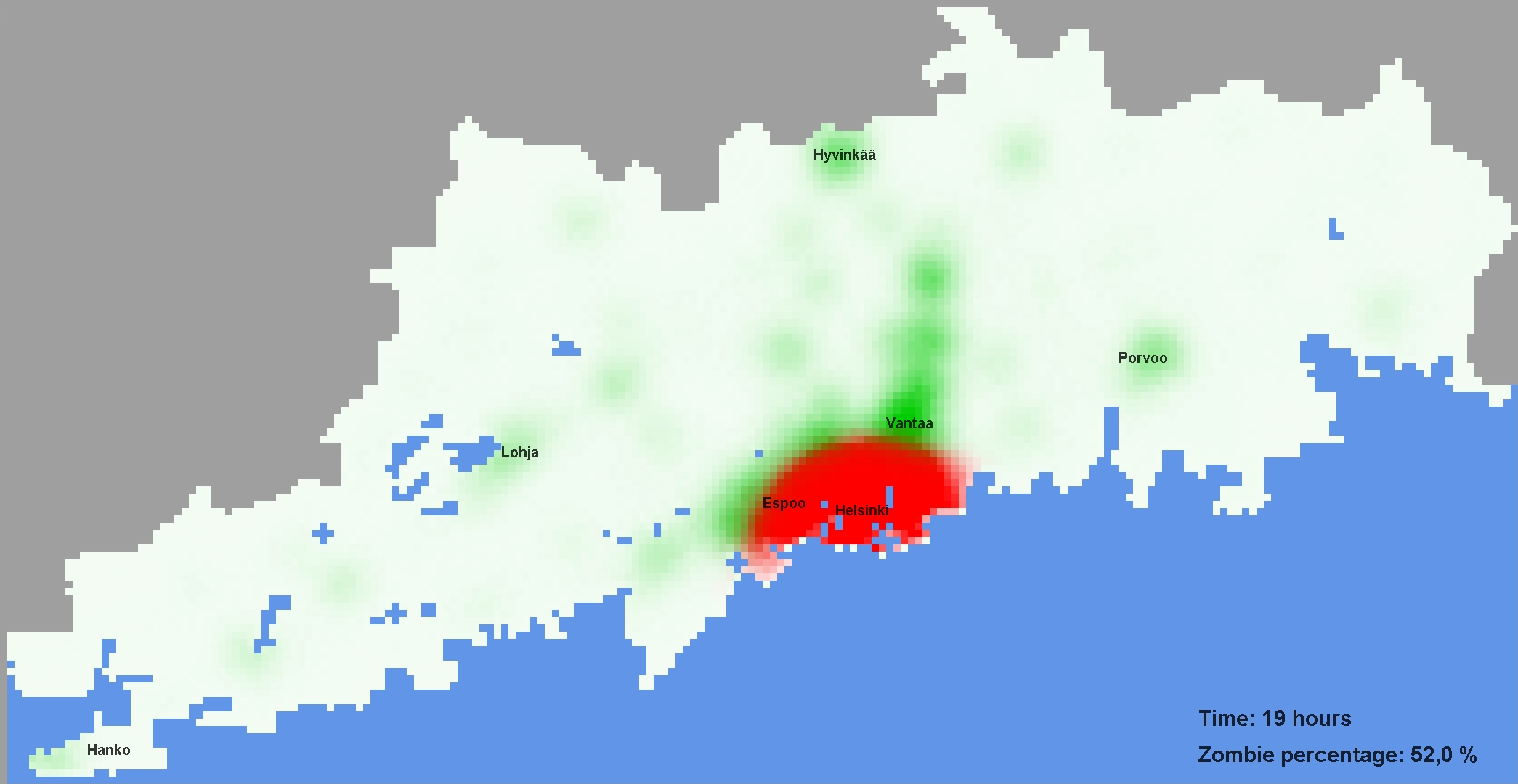}
	\end{subfigure}~
	\begin{subfigure}{.49\linewidth}
		\centering
		\includegraphics[width=\linewidth]{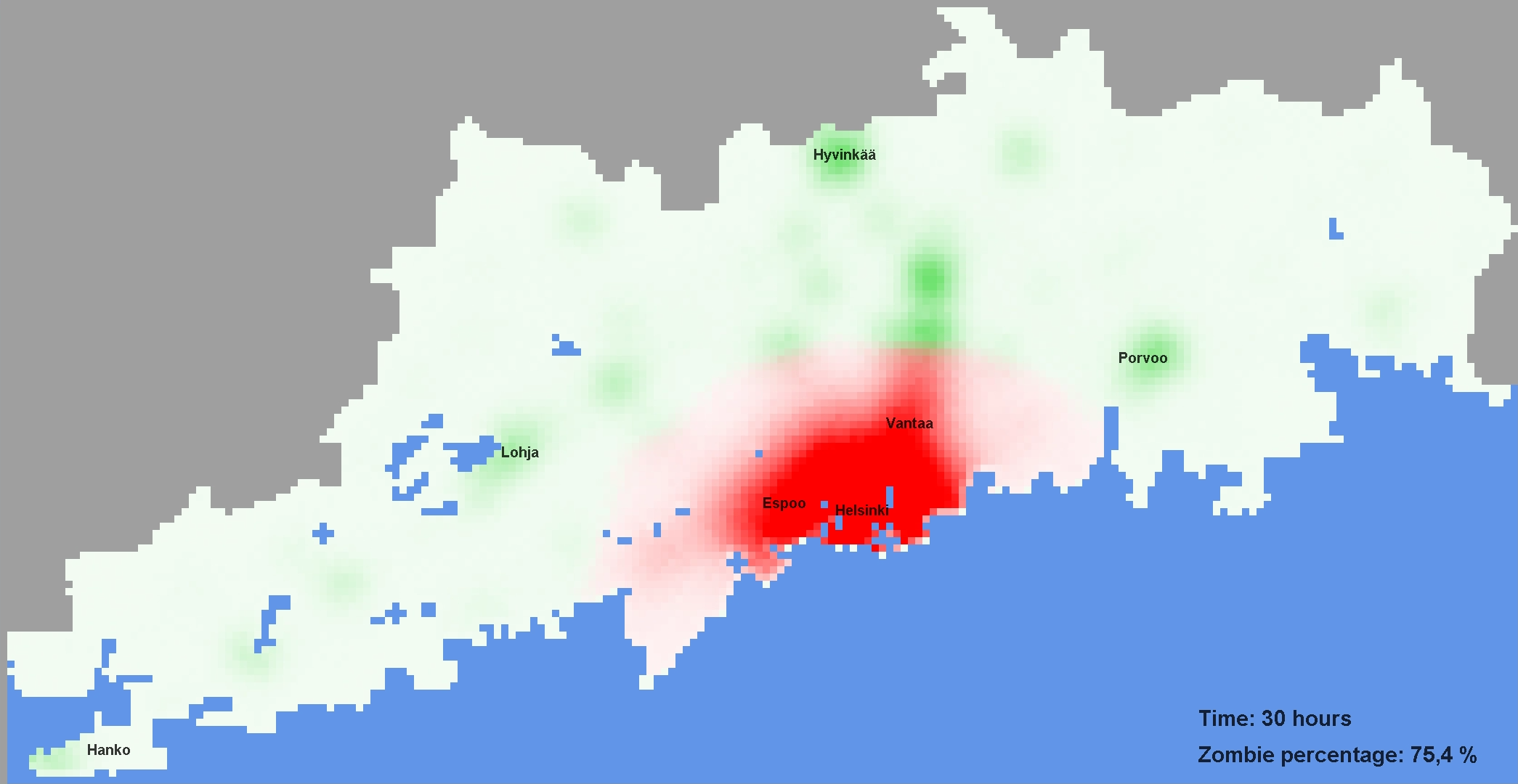}
	\end{subfigure}\\[1mm]
	\begin{subfigure}{.49\linewidth}
		\centering
		\includegraphics[width=\linewidth]{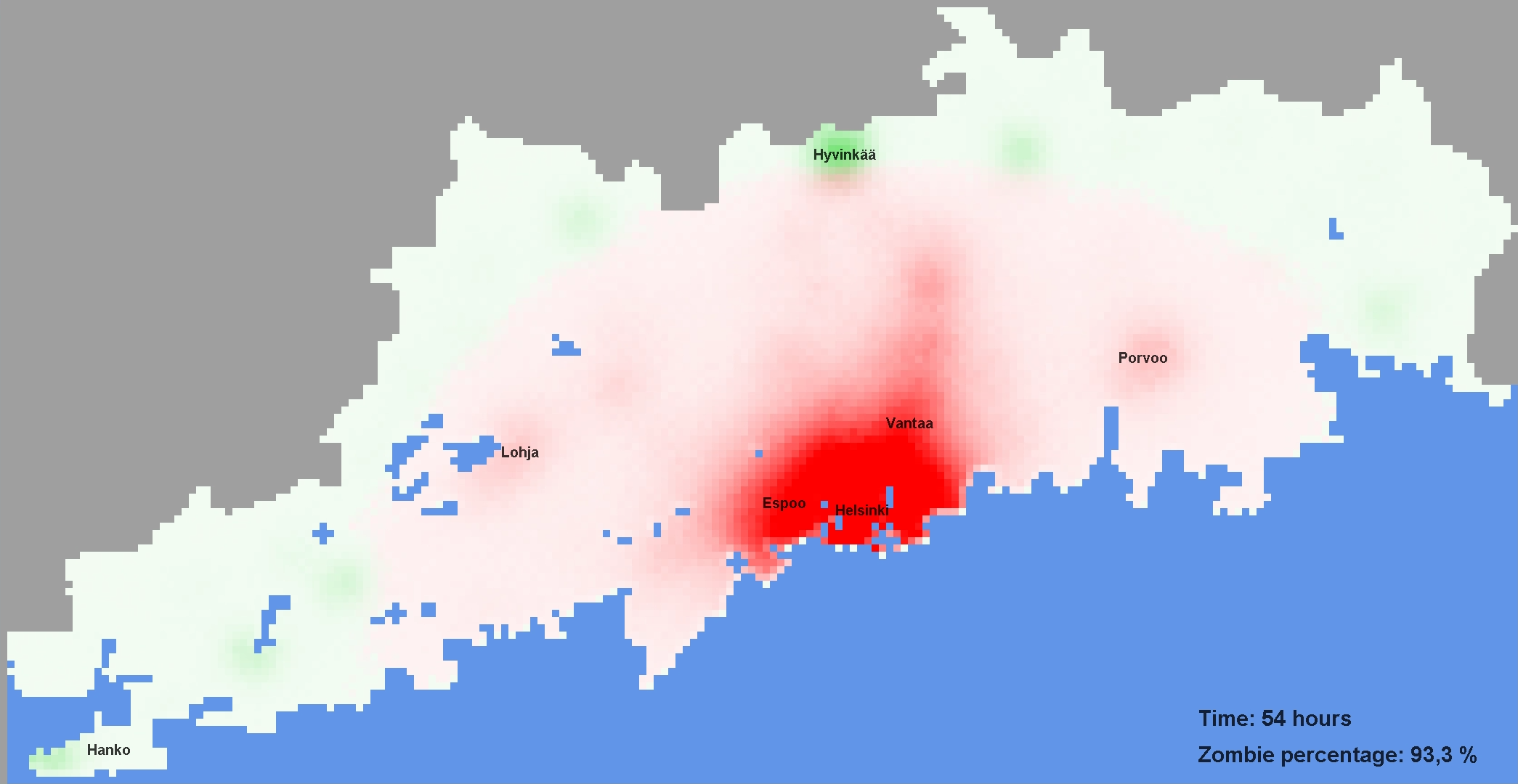}
	\end{subfigure}~
	\begin{subfigure}{.49\linewidth}
		\centering
		\includegraphics[width=\linewidth]{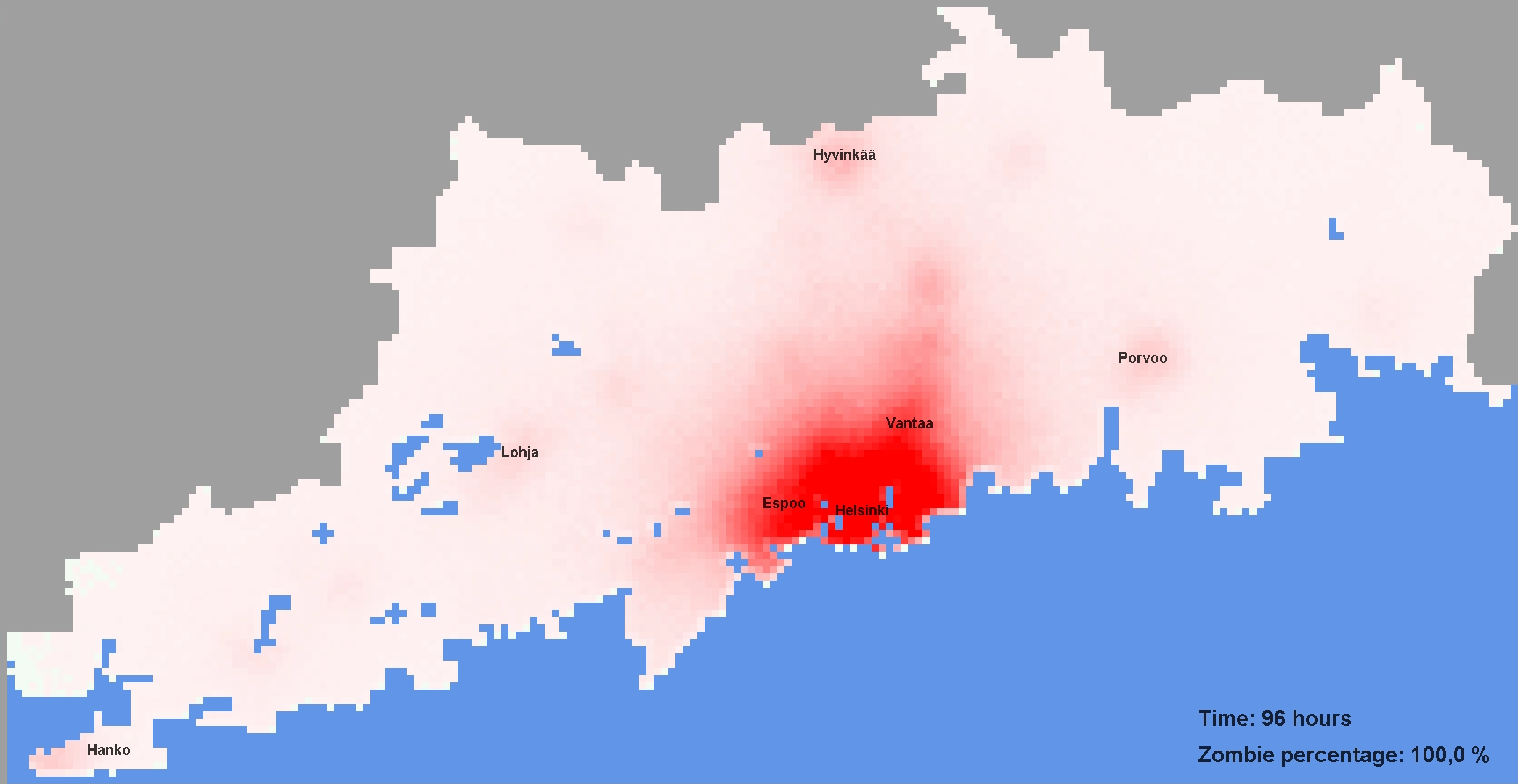}
	\end{subfigure}%
	\caption{Snapshots of an example simulation run in Scenario 1 at six different time steps, highlighting the spread of the epidemic. Humans are shown
in green and zombies in red. }
	\label{fig:simulation-example}
\end{figure}

Figure~\ref{fig:simulation-example} presents six snapshots from an example
simulation run in Scenario 1. Map animated videos demonstrating the progression
of single zombie epidemic runs under all three scenarios are available in the
project's GitHub repository \citep{repo}. The repository also includes Scala
codes for running the simulations and creating the animated videos.

\subsection{Scenario 1: No interventions}
\label{sec:results1}

\begin{figure}[!h]
	\centering
	\includegraphics[width=\linewidth]{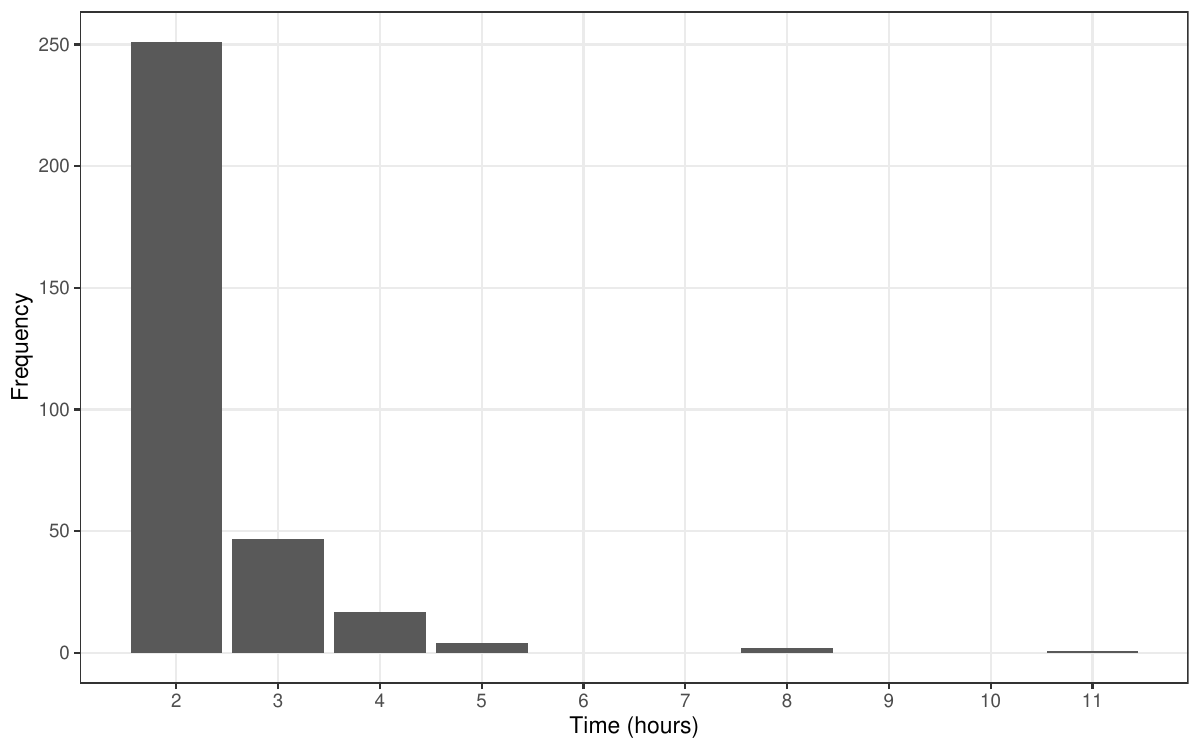}
	\caption{Hours taken for reaching a win in the runs won by humans in Scenario 1.}
	\label{fig:scenario-1-hours-to-win-humans}
\end{figure}

\begin{figure}[!h]
	\centering
	\includegraphics[width=\linewidth]{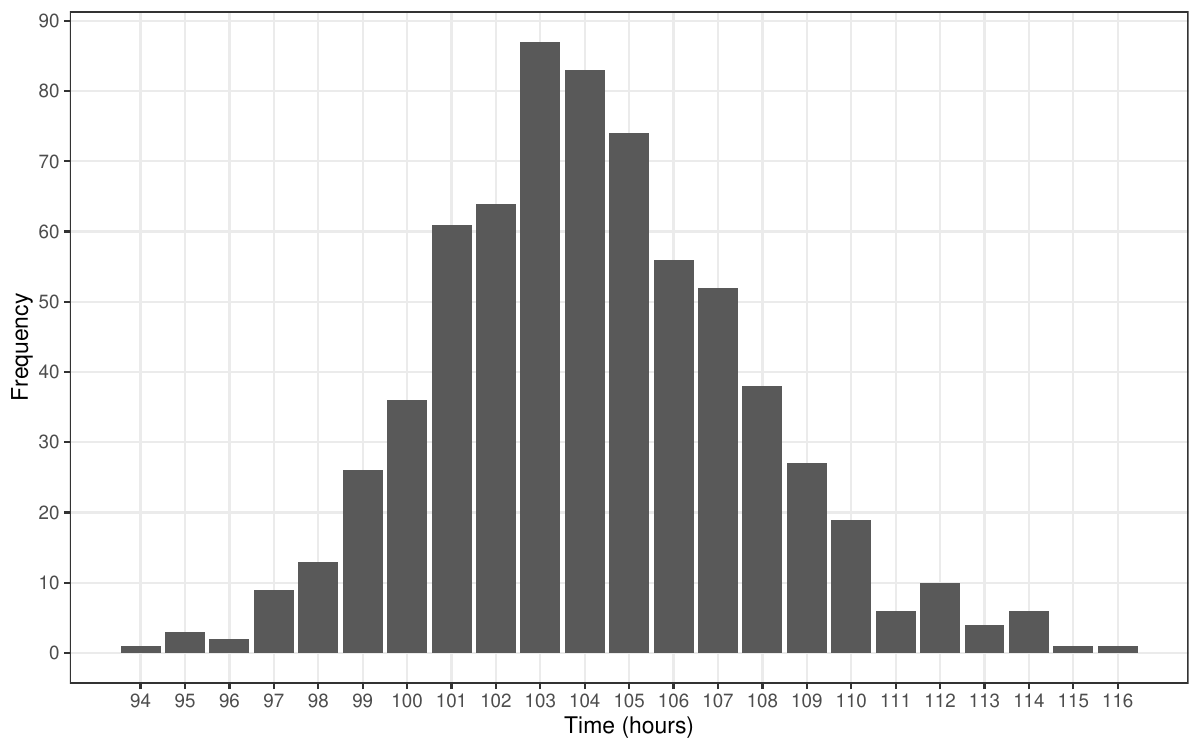}
	\caption{Hours taken for reaching a win in the runs won by zombies in Scenario 1.}
	\label{fig:scenario-1-hours-to-win-zombies}
\end{figure}

Figure~\ref{fig:scenario-1-hours-to-win-humans} and
Figure~\ref{fig:scenario-1-hours-to-win-zombies} present the time distributions
for human and zombie victories, respectively, in the scenario without
interventions. Majority of the human wins are achieved on the first few time
steps, when every zombie has by chance been defeated after only a few
interactions. Across all the simulation runs won by the humans, the number of
zombies at any given time point never exceeds four. For zombie victories, the
distribution of the time step of the wins is relatively symmetric, with the
majority of the wins occurring between the time steps 99 and 110. For zombies,
the time taken to achieve a win is mainly dictated by how long it takes for the
randomly walking horde of zombies to reach the map's edges, infecting all
encountered humans along the way.

\begin{figure}[!h]
	\centering
	\includegraphics[width=\linewidth]{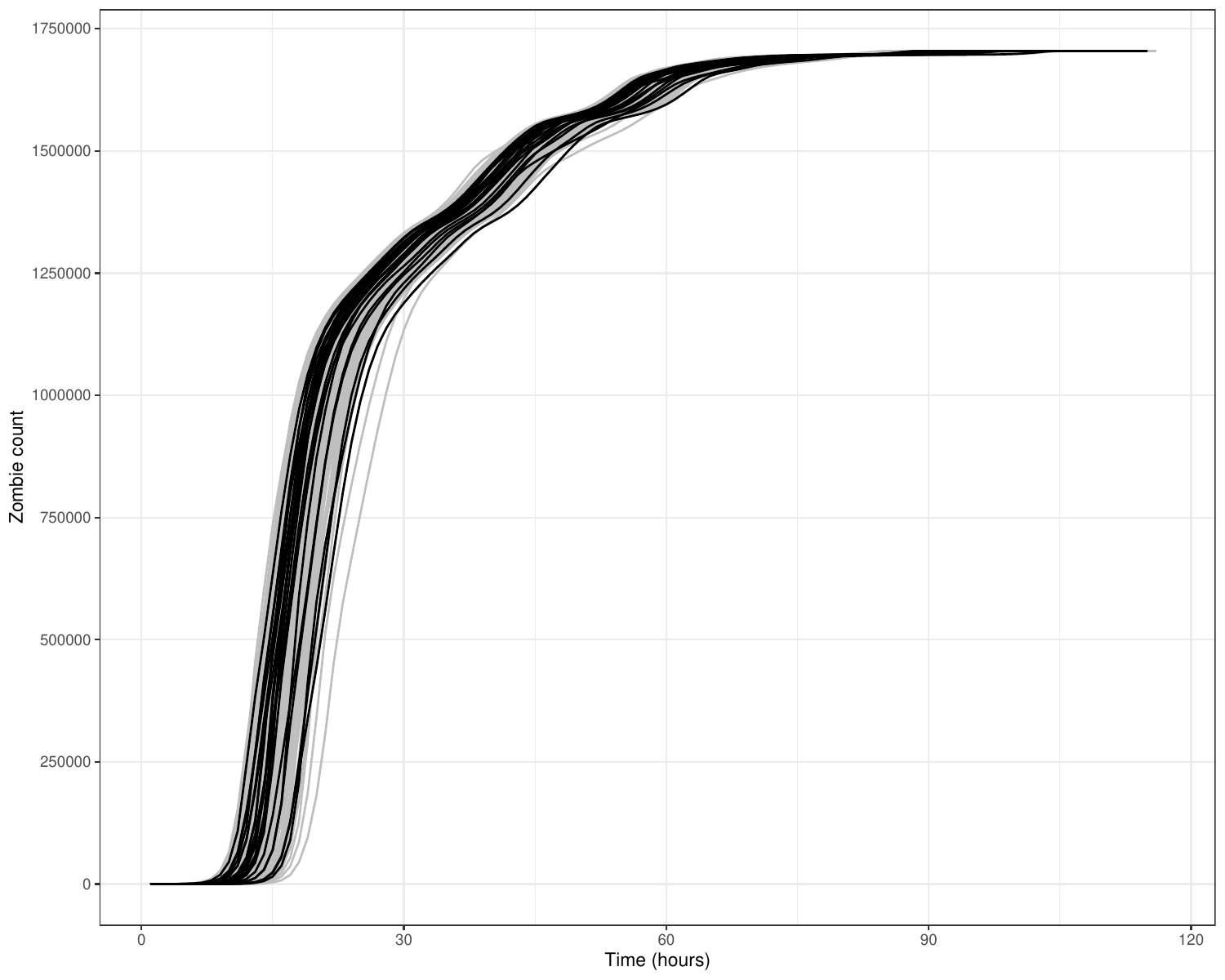}
	\caption{Development of the number of zombies over time in all runs of Scenario 1, with randomly chosen runs highlighted in black.}
	\label{fig:scenario-1-all-runs}
\end{figure}

Figure~\ref{fig:scenario-1-all-runs} illustrates the development of the total
number of zombies over time across all the runs of Scenario 1. Notably, runs
that end with a zombie victory show very similar patterns. There is only some
variability due to differences in when a critical mass of zombies is achieved.
Once this initial surge occurs, the rest of the run follows a similar trend.

\begin{figure}[!h]
	\centering
	\includegraphics[width=\linewidth]{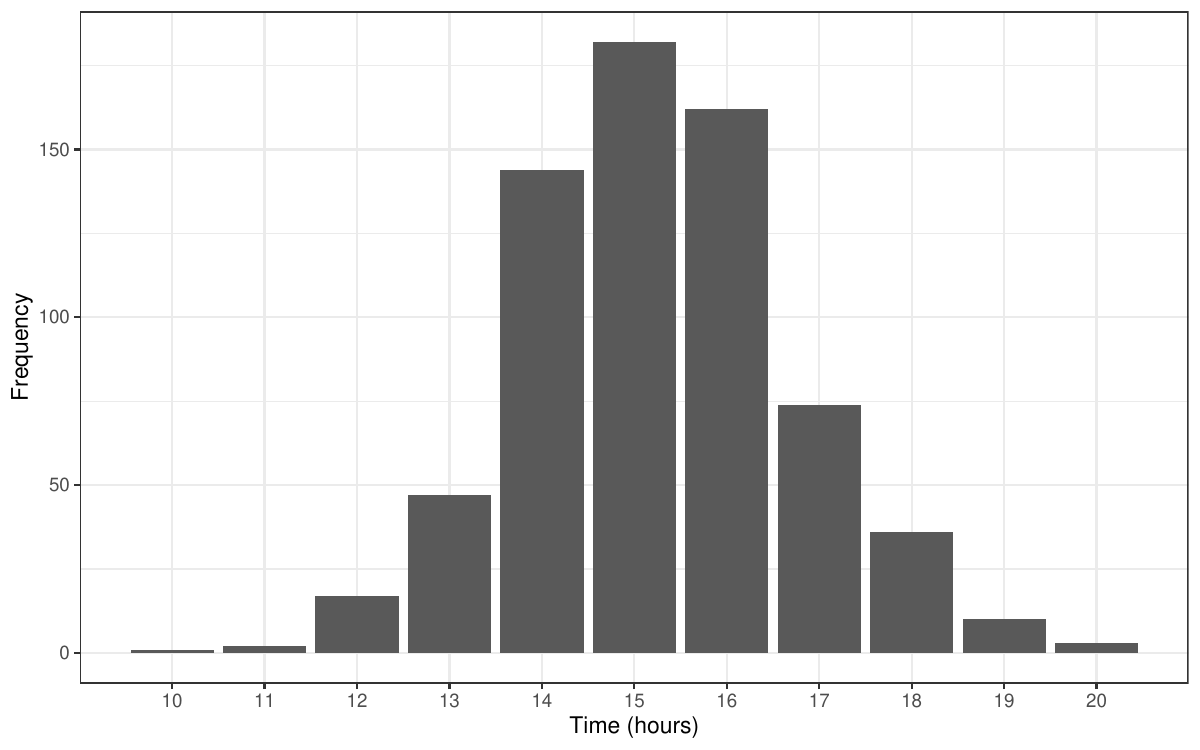}
	\caption{Hours taken for the first zombie to reach the border of the Helsinki region (Helsinki, Vantaa and Espoo) in Scenario 1.}
	\label{fig:scenario-1-edge-of-helsinki-time-distribution}
\end{figure}

Figure~\ref{fig:scenario-1-edge-of-helsinki-time-distribution} illustrates the
distribution of arrival times at which the first zombie reached the border of
the Helsinki region for all runs won by the zombies in Scenario 1. The primary
motivation for delaying the establishment of the quarantine zone was to account
for the time required for decision-making and physical preparations. This
distribution of arrival times was used to fine-tune the intervention delay in
both Scenario 2 and Scenario 3.

\subsection{Scenario 2: Strict quarantine region}
\label{sec:results2}

\begin{figure}[!h]
	\centering
	\includegraphics[width=\linewidth]{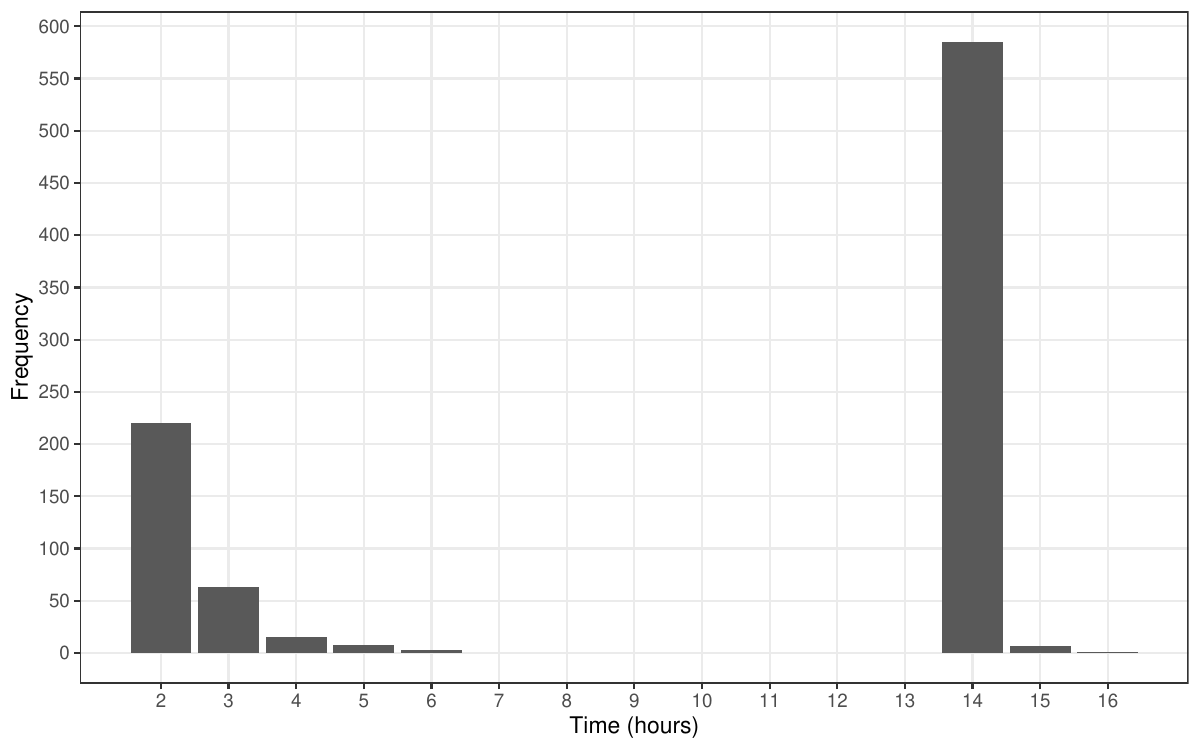}
	\caption{Hours taken for reaching a win in the runs won by humans in Scenario 2.}
	\label{fig:scenario-2-hours-to-win-humans}
\end{figure}

\begin{figure}[!h]
	\centering
	\includegraphics[width=\linewidth]{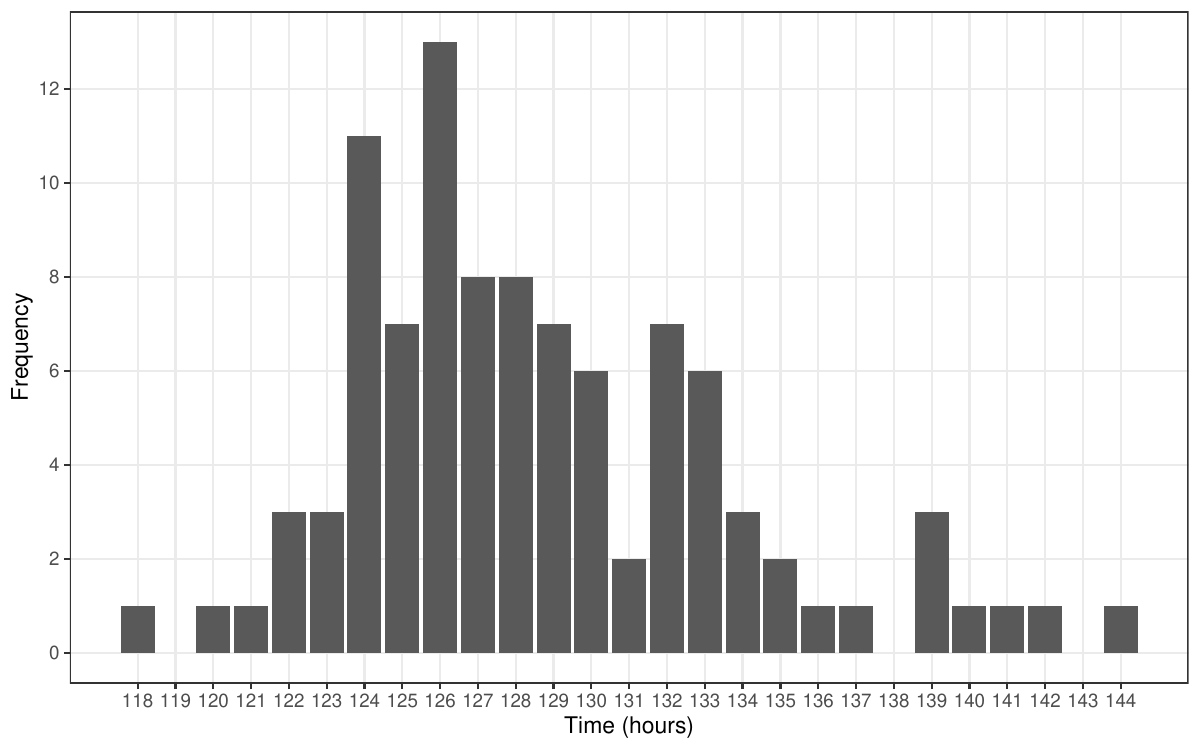}
	\caption{Hours taken for reaching a win in the runs won by zombies in Scenario 2.}
	\label{fig:scenario-2-hours-to-win-zombies}
\end{figure}

Figure~\ref{fig:scenario-2-hours-to-win-humans} and
Figure~\ref{fig:scenario-2-hours-to-win-zombies} show the distributions of time
taken for humans and zombies to achieve a victory, respectively, in Scenario 2.
In Scenario 2, human victories that occur early (before hour $6$) follow a
similar pattern to that in Scenario 1 (shown in
Figure~\ref{fig:scenario-1-hours-to-win-humans}). However, Scenario 2 also shows
a significant increase in human victories at time step 14, coinciding with the
timing of the intervention. A simulation run was classified as a human win if
the strict intervention had been implemented and there were no zombies outside
of the quarantined Helsinki region. In some of these simulations, where humans
won at time steps 15, or 16, a few zombies had managed to escape the Helsinki
region before the intervention began, but the zombies were quickly defeated by
humans, preventing further spread of the epidemic.

The distribution of the time step of the zombie victories in Scenario 2 shows
both, a higher average and a greater variance, compared to Scenario 1, as
illustrated in Figure~\ref{fig:scenario-2-hours-to-win-zombies}. Additionally,
the distribution in Scenario 2 seems to have more mass on the right tail.

\begin{figure}[!h]
	\centering
	\includegraphics[width=\linewidth]{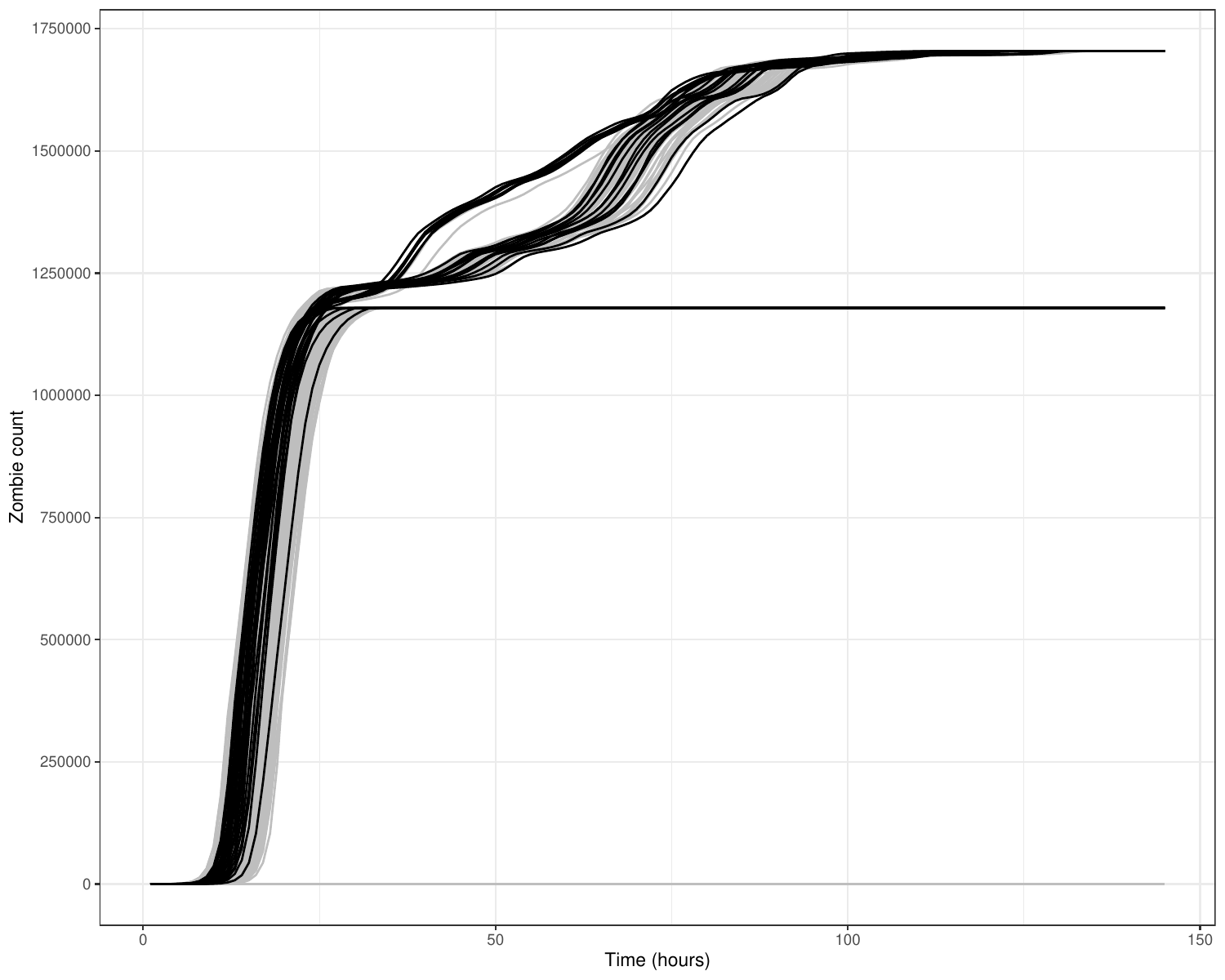}
	\caption{Development of the number of zombies over time in all runs of Scenario 2, with randomly chosen runs highlighted in black.}
	\label{fig:scenario-2-all-runs}
\end{figure}

Figure~\ref{fig:scenario-2-all-runs} illustrates the progression of the total
number of zombies over time for all runs of Scenario 2. In scenarios where
humans win due to the strict intervention, the entire Helsinki region (with a
population of approximately 1.2 million) still falls to the zombies. The zombies
within the Helsinki region are eventually defeated when they reach the
quarantine area border. Among the runs where zombies win, we can detect two
different groups. The differences between these two groups can be explained by
the location and the number of zombies that escaped the Helsinki region before
the strict quarantine was implemented.

\subsection{Scenario 3: An almost-strict quarantine region}
\label{sec:results3}

\begin{figure}[!h]
	\centering
	\includegraphics[width=\linewidth]{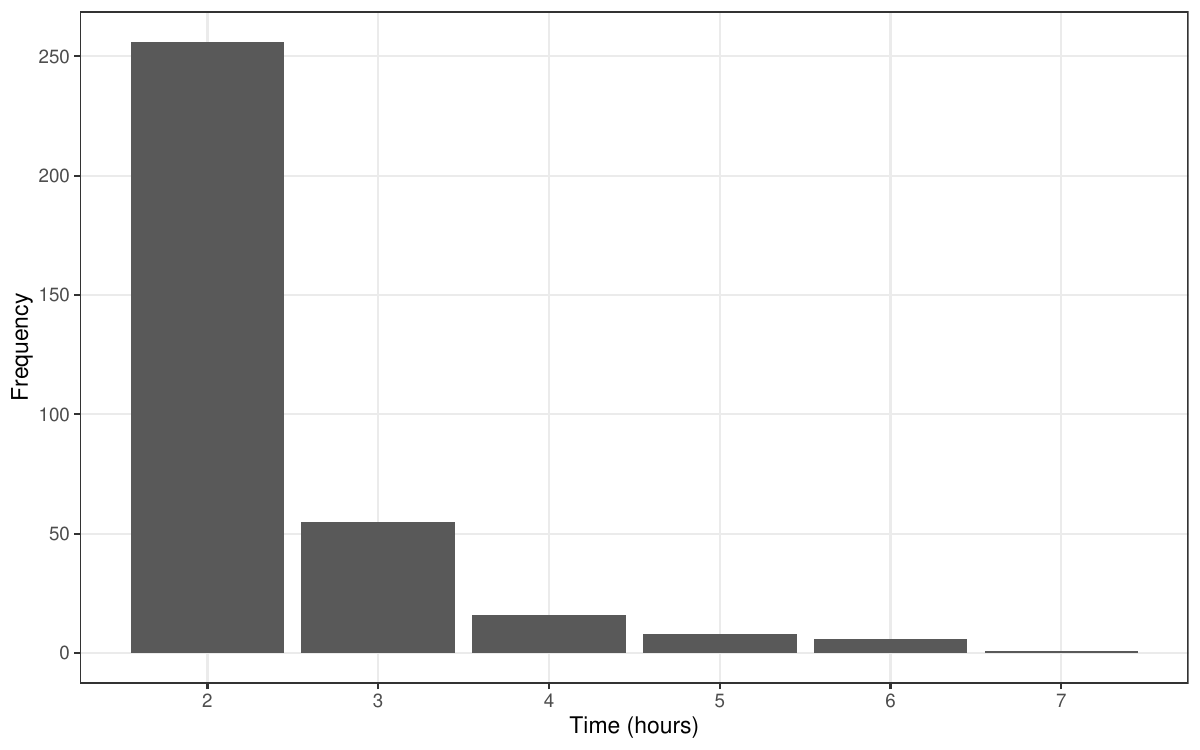}
	\caption{Hours taken for reaching a win in the runs won by humans in Scenario 3.}
	\label{fig:scenario-3-hours-to-win-humans}
\end{figure}

\begin{figure}[!h]
	\centering
	\includegraphics[width=\linewidth]{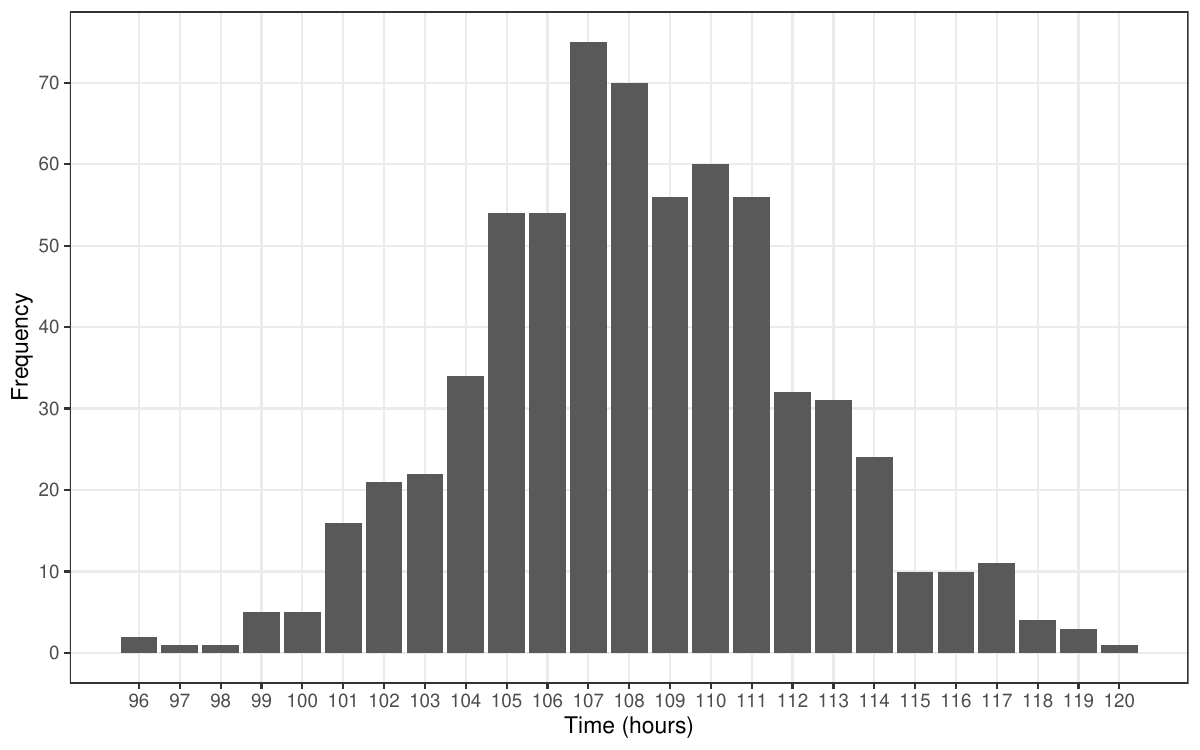}
	\caption{Hours taken for reaching a win in the runs won by zombies in Scenario 3.}
	\label{fig:scenario-3-hours-to-win-zombies}
\end{figure}

Figure~\ref{fig:scenario-3-hours-to-win-humans} and
Figure~\ref{fig:scenario-3-hours-to-win-zombies} show the distributions of time
taken for humans and zombies to achieve a victory, respectively, in Scenario 3. 

\begin{figure}[!h]
	\centering
	\includegraphics[width=\linewidth]{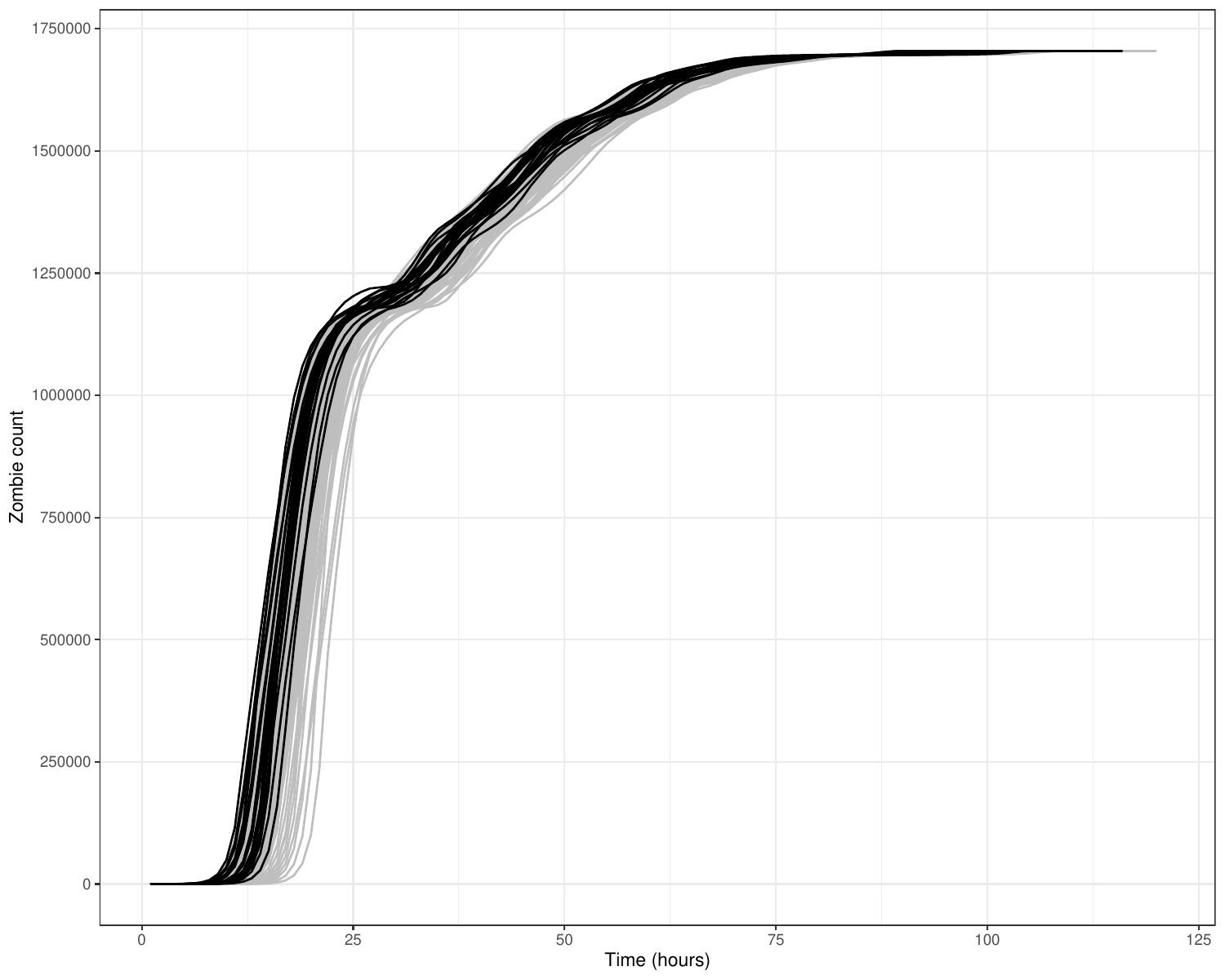}
	\caption{Development of the number of zombies over time in all runs of Scenario 3, with randomly chosen runs highlighted in black.}
	\label{fig:scenario-3-all-runs}
\end{figure}

As in Scenario~1 and Scenario~2, most of the variability in the development of
the total number of zombies over time in Scenario 3 (see
Figure~\ref{fig:scenario-3-all-runs}) is explained by differences in the timing
at which a critical mass of zombies is initially reached. However, the shape of
the curves  differ slightly from those in Scenario 1 (see Figure
\ref{fig:scenario-1-all-runs}), with a temporary slowdown in zombie growth
around time step $25$. This delay is attributed to the almost-strict
intervention; as the initial wave of zombies reaches the quarantine border and
is stopped with a high probability, the spread of the zombie epidemic is
temporarily slowed down until more and more zombies arrive and some of them
manage to cross the border.

\section{Discussion}
\label{sec:discussion}

Under the modeled human behavior, defense strategies (no, partial, or strict
intervention), and zombie profile, the simulation results clearly indicate that
only timely implemented strict measures can effectively defeat the zombies. The
majority of human victories occur when all the zombies are eliminated after only
a few interactions during the first few time steps of the simulation. If the
zombies survive and no measures are introduced at the beginning (time step 14 in
the simulations) of the zombie attack, even a single zombie can lead to an
uncontrollable epidemic. Partial measures preventing zombie's attempts to spread
outside of the quarantined area are effective in delaying the spread, but not in
stopping it. Interventions are effective in stopping an epidemic completely only
if they are strict and implemented without delay.

The main findings of this study emphasize the importance of timely and organized
responses as key to changing the course of  epidemics and saving lives. Rapid
and coordinated action is crucial for controlling and reversing disease spread.
In particular, implementing a stringent quarantine early on proves to be a
highly effective strategy. This study highlights the need for further
examination and simulation of extraordinary and severe events such as pandemics
to provide insights and guide preemptive preparations. Moreover, this study
highlights the value of public readiness for exceptional situations and the
importance of cooperation between authorities.

\section{Future prospects and concluding remarks}
\label{sec:future}

In this work, we utilized research on human behavior, common zombie culture, and
real geographic and population density data, in modeling zombie epidemics. We
simulated scenarios that offer insights into conditions similar to the recent
COVID-19 pandemic. Model parameter values, such as the incubation period and
human reaction probabilities, can be easily modified. Sample size, local
population densities and the movement of the agents can be altered.
Additionally, our model could be adapted to other geographical regions and to
more complex human behavior.

\cite{provitolo2011emergent}, for example, study the emergent human behavior
during crises, disasters and catastrophes by examining the different stages of
the disaster event (pre-, during, and post-event). For instance, human behavior
just before a disaster is influenced by factors such as the presence or absence
of warnings and the perceived imminence of the threat. Our model could be
expanded also to incorporate time-dependent behavior and to simulate the effects
of successful guidance from authorities. In addition to the human reactions
described in Section~\ref{sec:behavior}, \cite{provitolo2011emergent} identify
several additional behavior types. Our model could be further developed to
include, e.g., grouping (typical for human beings as it can increase survival
chances by increasing the odds of a successful defensive attack), sheltering
(another potentially efficacious method of increasing the probability of
survival), and altruistic suicide (of an individual, after being bitten, when in
the vicinity of loved ones).

\paragraph*{Acknowledgements.} The authors gratefully acknowledge support from
the Academy of Finland, decision number 346308 (Center of Excellence in
Randomness and Structures).

\bibliography{library}
\end{document}